\documentclass[preprint]{elsarticle}
\usepackage{graphicx}
\usepackage{dcolumn}
\usepackage{bm}
\usepackage{amsmath}
\begin{document}

\def\vri{\vec{r}_{i}}
\def\vrj{\vec{r}_{j}}
\def\rij{r_{ij}}
\def\vrij{\vec{r}_{ij}}
\def\drij{\hat{r}_{ij}}
\def\vdr{\delta\vec{r}}
\def\dr{\delta{r}}
\def\s{\hat{s}}

\title{Jamming I: A volume function for jammed matter }

\author[focal]{Chaoming Song}
\author[focal]{Ping Wang}
\author[focal]{Yuliang Jin}
\author[focal]{Hern\'an A. Makse\corref{cor1}}
\ead{hmakse@lev.ccny.cuny.edu}
\cortext[cor1]{Corresponding author}
\address[focal]{Levich Institute and Physics Department, City College of New
  York, New York, NY, USA 10031}

\date{\today }
\begin{abstract} {\bf We introduce a ``Hamiltonian''-like function,
    called the volume function, indispensable to describe the ensemble
    of jammed matter such as granular materials and emulsions from a
    geometrical point of view.  The volume function represents the
    available volume of each particle in the jammed systems.  At the
    microscopic level, we show that the volume function is the Voronoi
    volume associated to each particle and in turn we provide an
    analytical formula for the Voronoi volume in terms of the contact
    network, valid for any dimension. We then develop a statistical
    theory for the probability distribution of the volumes in 3d to
    calculate an average volume function coarse-grained at a
    mesoscopic level. The salient result is the discovery of a
    mesoscopic volume function inversely proportional to the
    coordination number.  Our analysis is the first step toward the
    calculation of macroscopic observables and equations of state
    using the statistical mechanics of jammed matter, when
    supplemented by the condition of mechanical equilibrium of jamming
    that properly defines jammed matter at the ensemble level.}
\end{abstract}




\maketitle

\section{Introduction}

The development of a statistical mechanics for granular matter and
other jammed materials presents many difficulties.  First, the
macroscopic size of the constitutive particles forbids equilibrium
thermalization of the system. Second, the fact that energy is
constantly dissipated via frictional interparticle forces further
renders the problem outside the realm of equilibrium statistical
mechanics.
In the absence of energy conservation laws, a new statistical
approach is needed in order to describe the system properties.
Along this line of research, Edwards \cite{sirsam} has proposed to
replace the energy by the volume as the conservative quantity of
the system. Then a canonical partition function of jammed states
can be defined and a statistical mechanical analysis ensues.

While it is always possible to measure the total volume of the
system, it is unclear how to treat the volume fluctuations at the
microscopic level.  Thus, it is still an open problem the definition
of an analogous ``Hamiltonian'' that describes the microstates of
jammed matter.
Such a ``Hamiltonian'' is called the volume function
\cite{sirsam}, denoted ${\cal W}$.
The idea is to partition the granular material into $N$ elements
and associate an additive volume function to them, ${\cal W}_i$,
such that the total system volume, ${\cal W}$, is
\begin{equation}
{\cal W} = \sum_{i=1}^N {\cal W}_i.
\end{equation}

From a theoretical perspective, initial attempts to define the
volume function involved modelling under mean-field approximations
proposed by Edwards \cite{sirsam} not taking into account the
contact network.  The necessary definition in terms of the
internal degrees of freedom has been pursued by Ball and
Blumenfeld \cite{ball,new0} who have shown by an exact triangulation
method that the volume defining each grain can be given in terms
of the contact points using vectors constructed from them.  The
method consists of defining shortest loops of grains in contact
with one another, thus defining the void space around a central
grain.

A simpler version for the volume function was also given by Edwards as
the area in 2d or volume in 3d encompassing the first coordination
shell of the grains in contact \cite{edwards-tensorial}.  The
resulting volume is the antisymmetric part of the fabric tensor, the
significance of which is its appearance in the calculation of stress
transmission through granular packings \cite{ball}.
This definition is only an approximation of the space available to
each grain since there is an overlap of volumes for grains
belonging to the same coordination shell. Thus, it overestimates
the total volume of the system: $\sum {\cal W}_i > {\cal W}$
\cite{edwards-tensorial}.

Furthermore, both definitions of ${\cal W}_i$ in
\cite{ball,new0,edwards-tensorial}
are proportional to the coordination number of the grain.  This is
in contrast to expectation since the free volume available to a
grain should decrease as the number of contacts increase. Indeed,
this observation is corroborated by experimental studies of jammed
granular matter using X-ray tomography \cite{aste} to determine
the volume per grain versus coordination number.


Based on the idea that the volume function represents a free
volume available per grain, we introduce a new ``Hamiltonian'' to
describe the microstates of jammed matter. We analytically
calculate the volume function and demonstrate that it is equal to
the Voronoi volume associated to each particle, partitioning the
space into a set of regions, associating all grain centroids in
each region to the closest grain centroid.  Even though the
Voronoi construction successfully tiles the system, its drawback
in its use as a volume function was that, so far, there was no
analytical formula to calculate it. Our approach provides this
formula
in terms of the contact network.  Furthermore we introduce a
theory of volume fluctuations to calculate a coarse-grained
average volume function defined at the mesoscopic level that
reduces the degrees of freedom to only the coordination number
$z$.  We find that the volume function is inversely proportional
to $z$ in agreement with the tomography experiments of
\cite{aste}.  Our analysis also provides an equation of state,
relating volume with coordination number in the limit of fully
random system. Indeed, it predicts with good accuracy the limiting
cases of random loose and random close packing fractions. Our
results allow construction of a statistical partition function
from which macroscopic observables can be calculated.  This case
is treated in more detail in the second part of this work: Jamming II
\cite{jamming2}.

\subsection{Outline}

This paper is the first installment in a series of papers devoted to
different aspects of jammed matter and is the main theoretical
contribution for the subsequent statistical mechanics approach. The
present paper is an extended version of the Supplementary Information
Section I in \cite{jamming2}. The outline is as follows: Section
\ref{hamiltonian} details the development of the microscopic volume
function in terms of the particle coordinates. The relation of this
form with the Voronoi volume of each particle is discussed in Section
\ref{voronoi-formula}. Section \ref{proba} discusses the statistical
theory to calculate the probability distribution of the Voronoi
volumes leading to the average mesoscopic volume function discussed in
Section \ref{meso}. Section \ref{test} tests the assumptions of the
theory and we finish with the outlook in Section \ref{outlook}.




\section{Microscopic volume function}
\label{hamiltonian}

We start by defining a volume function for rigid spherical grains of
equal size in terms of particle positions.  The volume function
represents the available volume to the particle with the constraint of
fixed total system volume.  Since we are dealing with rigid jammed
particles, the available free volume is in principle zero since the
particle by definition cannot move.  However, we allow the particle to
move by introducing a soft interparticle potential and then taking the
limit of particle rigidity to infinity. The resulting volume is
well-defined, representing the free volume associated with each grain
in the jammed packing.

\begin{figure}
  \centering {
\resizebox{8cm}{!}{\includegraphics{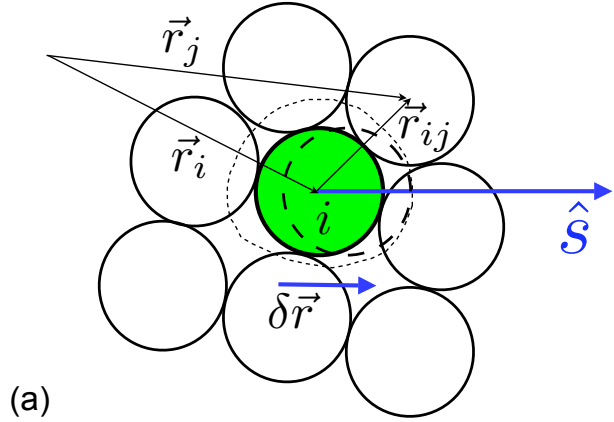}} }
  \caption{Definition of the volume function. Particle $i$ moves in
    the direction of $\hat s$ as indicated such that the deformation
    energy is below a threshold $\epsilon$. When averaged over all
    directions $\hat s$, this process defines the available volume
    under the threshold $\epsilon$ demarcated by the thin dotted line.
    The volume function is then obtained in the limit of infinite
    rigidity of the particles, $\alpha\to\infty$.}
\label{volumea}
\end{figure}

We consider a rigid particle of radius $R$ jammed at position
$\vri$ in contact with another particle at position $\vrj$ such
that $\vrij=\vrj-\vri = r_{ij} \hat{r}_{ij}$ (see Fig.
\ref{volumea}). In order to calculate the volume associated with
such a particle we allow it to move by introducing a generic
interparticle soft-potential, $f_\alpha$, determined by an exponent
$\alpha$ governing the rigidity of the particles (see below). A
small energy threshold $\epsilon$ is introduced in order to drive
the particle in a certain direction $\hat{s}$ as indicated in Fig.
\ref{volumea}.  A small displacement $\vdr = \delta r \s$ of
particle $i$ along the $\s$ direction results in an increase of
energy $E$ between particles $i$ and its neighbors:

\begin{equation} E = \sum_{\vdr\cdot\vrij > 0}f_\alpha(|\vrij|-|\vrij
+ \vdr|) \approx \sum_{\s\cdot\drij > 0}f_\alpha((\s\cdot\drij)\dr),
\label{E}
\end{equation}
where the sum is taken over all the neighbors of particle $i$. The
soft pair potential $f_\alpha$ can be any repulsive function
provided it approaches the hard sphere limit when the control
parameter $\alpha \to \infty$, implying
\begin{equation}
\lim_{\alpha\rightarrow\infty}\frac{f_\alpha(x)}{f_\alpha(y)} = 0,
\forall\ x < y.
\end{equation}
[A possible function is simply $f_\alpha(x) = x^\alpha$, the case
$\alpha=5/2$ corresponds to the Hertz potential].  This
condition implies that

\begin{equation}
\begin{split}
  \lim_{\alpha\rightarrow\infty}\sum_i f_\alpha(x_i) =& f_\infty(\max_i
  (x_i))\times\lim_{\alpha\rightarrow\infty}\sum_{i}\frac{f_\alpha(x_i)}
  {f_\infty(\max_i (x_i))} \\
  =& f_\infty(\max_i (x_i)),
\end{split}
\end{equation}
and therefore from Eq. (\ref{E}), we obtain:
\begin{equation}
  E=f_\infty(\max_i (\s\cdot\drij)\dr).
\label{energy}
\end{equation}

We define the available volume to a grain, ${\cal W}_i^{\rm a}$, under the energy
threshold $\epsilon$ as:
\begin{equation} {\cal W}_i^{\rm a} = \int \Theta(\epsilon - E) dV =
  \oint \int_0^\infty \Theta(\epsilon - E) \dr^{d-1} d[\dr] ds,
\end{equation}
where $d$ is the dimension of the system, $ds$ is an
infinitesimal solid angle in $d$ dimensions, and the available
volume ${\cal W}_i^{\rm a}$ is averaged over all the directions of
the $d$ dimensional solid angle.

The integration over $\dr$ can be simplified when
$\alpha\rightarrow\infty$ since for a fixed direction $\s$ we have:

\begin{equation}
\begin{split}
\lim_{\alpha\rightarrow\infty} \int_0^\infty \Theta(\epsilon - E)
\dr^{d-1} d[\dr] &\\
=\lim_{\alpha\rightarrow\infty} \int_0^\infty
\Theta[\epsilon - f_\alpha(\max_{\s\cdot\drij > 0}(\s\cdot&\drij)\dr)]
\dr^{d-1} d[\dr] \\
\propto &\min_{\s\cdot\drij > 0}(\s\cdot\drij)^{-d}.
\end{split}
\end{equation}
Thus, we obtain:

\begin{equation}
{\cal W}_i^{\rm a} \propto \oint \min_{\s\cdot\drij > 0}(\s\cdot\drij)^{-d} ds.
\label{va}
\end{equation}

Equation (\ref{va}) can be interpreted as follows: for each
direction $\s$, the available volume is determined by the particle
position whose projection of the distance to particle $i$ in the
$\s$ direction is minimal. The total volume is then the average
over all directions $\s$. The proportionality constant in Eq.
(\ref{va}) can be determined because ${\cal W}_i^{\rm a}$ is equal
to the volume of the grains, $V_g$, when the coordination number
$z\rightarrow\infty$, suggesting that, in this limit,
$\min_{\s\cdot\drij > 0}(\s\cdot\drij)^{-d} = 1$ for any $\s$.
That is,
\begin{equation}
{\cal W}_i^{\rm a} = \frac{V_g}{\oint
  ds}\oint \min_{\s\cdot\drij > 0}(\s\cdot\drij)^{-d} ds.
\end{equation}

For mono-disperse spherical particles, $V_g = \frac{R^d}{d}\oint
ds$. Thus, we have
\begin{equation}
\begin{split}
{\cal W}_i^{\rm a} = \frac{1}{d}\oint
  \left(\min_{\s\cdot\drij > 0}\frac{R}{\s\cdot\drij}\right)^{d} ds \\
  =\frac{1}{d}\oint \left(\min_{\s\cdot\drij >
      0}\frac{\rij}{2\s\cdot\drij}\right)^{d} ds,
\end{split}
\label{vi}
\end{equation}
where we have replaced $\rij = 2R$ for nearest neighbors in the
last equation. This allows us to generalize the volume formula to
satisfy additivity and relate it to the Voronoi volume as shown
below.

Equation (\ref{vi}) is not additive, and is different from
the Voronoi volume since only contacting particles are considered
in the calculation. Strictly speaking, only at the limit when the
coordination number $z\rightarrow\infty$, and all the geometrical
constraints come from contacting particles, the volume function
Eq. (\ref{vi}) converges to the Voronoi volume.
However, the formula becomes additive when considering all
particles rather than the nearest neighbors in the calculation of
the minimum integrand in Eq. (\ref{vi}).  This approach is
justified since non-contact particles may contribute to the energy
of deformation in Eq. (\ref{energy}). Further, if the minimum in
Eq. (\ref{vi}) is taken over all the particles in the packing,
${\cal W}_i^{\rm a}$ is exactly equal to the Voronoi volume, which
is obviously additive.  In turn we provide a formula for the
calculation of the Voronoi volume in terms of particle positions,
as shown next.

\section{A formula for the Voronoi volume}
\label{voronoi-formula}

First, we recall the definition of a Voronoi cell as a convex
polygon whose interior consists of all points which are closer to
a given particle than to any other.

\begin{figure}
\centering {
\resizebox{8cm}{!}{\includegraphics{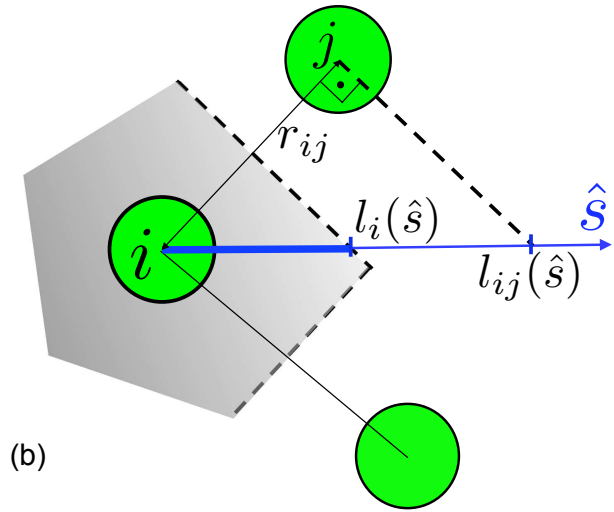}}
}
\caption{ The limit of the Voronoi cell (grey area) of particle
  $i$ in the direction $\hat s$ is $l_{i}(\hat{s}) =
  \min_j l_{ij}(\hat{s})/2$, where
$l_{ij} = r_{ij}/ \cos \theta_{ij}$. Then the Voronoi volume is
  proportional to the integration of $l_{i}(\hat{s})$
  over $\s$ as in Eq.  (\ref{vor1}).
 }
\label{volumeb}
\end{figure}

Formally, the volume of the Voronoi cell of particle $i$ can be
calculated as (see Fig. \ref{volumeb}):
\begin{equation} \label{vor0} {\cal W}_i^{\rm vor} = \oint
  \int_0^{l_i(\s)} r^{d-1}dr ds = \frac{1}{d} \oint l_i(\s)^d ds,
\end{equation}
where $l_i(\s)$ is the distance from particle $i$ to the boundary of
its Voronoi cell in the $\s$ direction. Note that this definition is
valid whether the particle $j$ is in contact with $i$ or not. If we
denote the distance from particle $i$ to any particle $j$ at $\s$
direction as
\begin{equation}
l_{ij}(\s)
\equiv \rij/(\s\cdot\drij),
\end{equation}
then $l_i(\s)$ is the minimum of $l_{ij}(\s)/2$ over all the particles
$j$ for any $l_{ij}(\s) > 0$ (see Fig. \ref{volumeb}). This leads to

\begin{equation}
l_i(\s) = \min_{l_{ij}(\s) > 0} l_{ij}(\s)/2 = \min_{\s\cdot\drij >
  0}\frac{\rij}{2\s\cdot\drij}.
\end{equation}
Substituting into Eq.  (\ref{vor0}), we prove that the Voronoi volume
is indeed the volume available per particle as calculated in
Eq. (\ref{vi}):
\begin{equation} \label{vor1}
{\cal W}_i^{\rm vor} = \frac{1}{d} \oint \left(\min_{\s\cdot\drij
> 0}(\frac{\rij}{2\s\cdot\drij})\right)^d ds = {\cal W}_i^{\rm a}.
\end{equation}

Formula (\ref{vor1}) can be rewritten as
\begin{equation}
{\cal W}_i^{\rm vor} =
\frac{1}{\oint ds} \oint {\cal W}_i^s ds = \langle {\cal W}_i^s
\rangle_s,
\end{equation}
where we define the orientational volume for the $i$ particle in the
$\s$ direction as:
\begin{equation}
  {\cal W}_i^s \equiv V_g \left(\frac{1}{2R} \min_{\s\cdot\drij > 0}
    \frac{\rij}{\s\cdot\drij}\right) ^ d.
\label{voronoi_s}
\end{equation}

Let us recapitulate and recall the three volumes defined so far which
are interrelated: the Voronoi and the available volume which satisfy
${\cal W}_i^{\rm vor} = {\cal W}_i^{\rm a}$, and the orientational
volume which satisfies $\langle {\cal W}_i^{s} \rangle_s = {\cal
  W}_i^{\rm vor} $. All the quantities are additive, thus they provide
the total volume of the system:
\begin{equation} {\cal W}= \sum_i {\cal W}_i^{\rm vor} = \sum_i {\cal
    W}_i^{\rm a} = \sum_i \langle {\cal W}_i^s \rangle_s.
\end{equation}
For isotropic packings, ${\cal W}_i^s$ (without the average over $\s$)
is also additive since the choice of orientation $\s$ is
arbitrary. Thus, we obtain:
\begin{equation}
  {\cal W} = \sum_i \langle {\cal W}_i^s \rangle_s = \langle \sum_i
  {\cal W}_i^s \rangle_s = \sum_i {\cal W}_i^s .
\end{equation}
This property reduces the calculations, since there is no need for an
orientational average.  We define the orientational free volume
function as:

\begin{equation}
  W={\cal
    W}_i^s - V_g,
\end{equation}
and the reduced orientational free volume function
$${w_i}^s = \frac{{\cal W}_i^s -
  V_g}{V_g},$$
 with its average value, $w$, over the particles, $i$,
for isotropic systems as:

\begin{equation}
  w \equiv \langle w_i^s\rangle_i \equiv
  \frac{\langle {\cal W}_i^s \rangle_i- V_g}{V_g} = \left\langle \left(\frac{1}{2R}
  \min_{\s\cdot\drij
    > 0}\frac{\rij}{\s\cdot\drij}\right)^d \right\rangle_i -1.
\label{phi}
\end{equation}
This average orientational volume function requires only averaging
over the particles $i$ but not over $\s$.  The more general form of
Eq.  (\ref{vor1}) allows study of anisotropic systems, a case left to
future work. We notice that the average of the orientational volume
over the particles (for a fixed $\s$) is equal to the average of the
Voronoi volume over the particles: $\langle {\cal W}_i^s \rangle_i =
\langle {\cal W}_i^{\rm vor} \rangle_i$.

The relevance of Eq. (\ref{vor1}) is that:

{\it(i)} It provides a formula for the Voronoi volume for any
dimension in terms of particle positions $\vri$.

{\it (ii)} It suggests that the Voronoi volume is the volume
function of the system since it is indeed ${\cal W}_i^{\rm a}$ as
calculated from an independent point of view.

{\it (iii)} It allows for the calculation of macroscopic
observables via statistical mechanics.

However, further analytical developments are difficult since:

{\it (i)} The volume function is a complicated non-local, not
pair-wise function of the coordinates.

{\it (ii)} It requires the use of field-theoretical methods.

{\it (iii)}
It cannot be factorized into a single particle partition function,
implying that there are intrinsic strong correlations in the
system. Such correlations are implicit in the global minimization
in Eq. (\ref{vor1}) which, in practice, is restricted to a few
coordination shells and defines a mesoscopic Voronoi length scale,
which will be of use below.

To circumvent the above difficulties, we present a theory of volume
fluctuations to coarse grain ${\cal W}_i^{\rm vor}$ over this
mesoscopic length scale. This coarsening reduces the degrees of
freedom to one variable, the coordination number of the grains, and
calculates the average mesoscopic volume function Eq. (\ref{phi})
amenable to statistical calculations.

\section{The probability distribution of Voronoi volumes}
\label{proba}

\begin{figure}
  \centering { \vbox {
\resizebox{8cm}{!}
      { \includegraphics{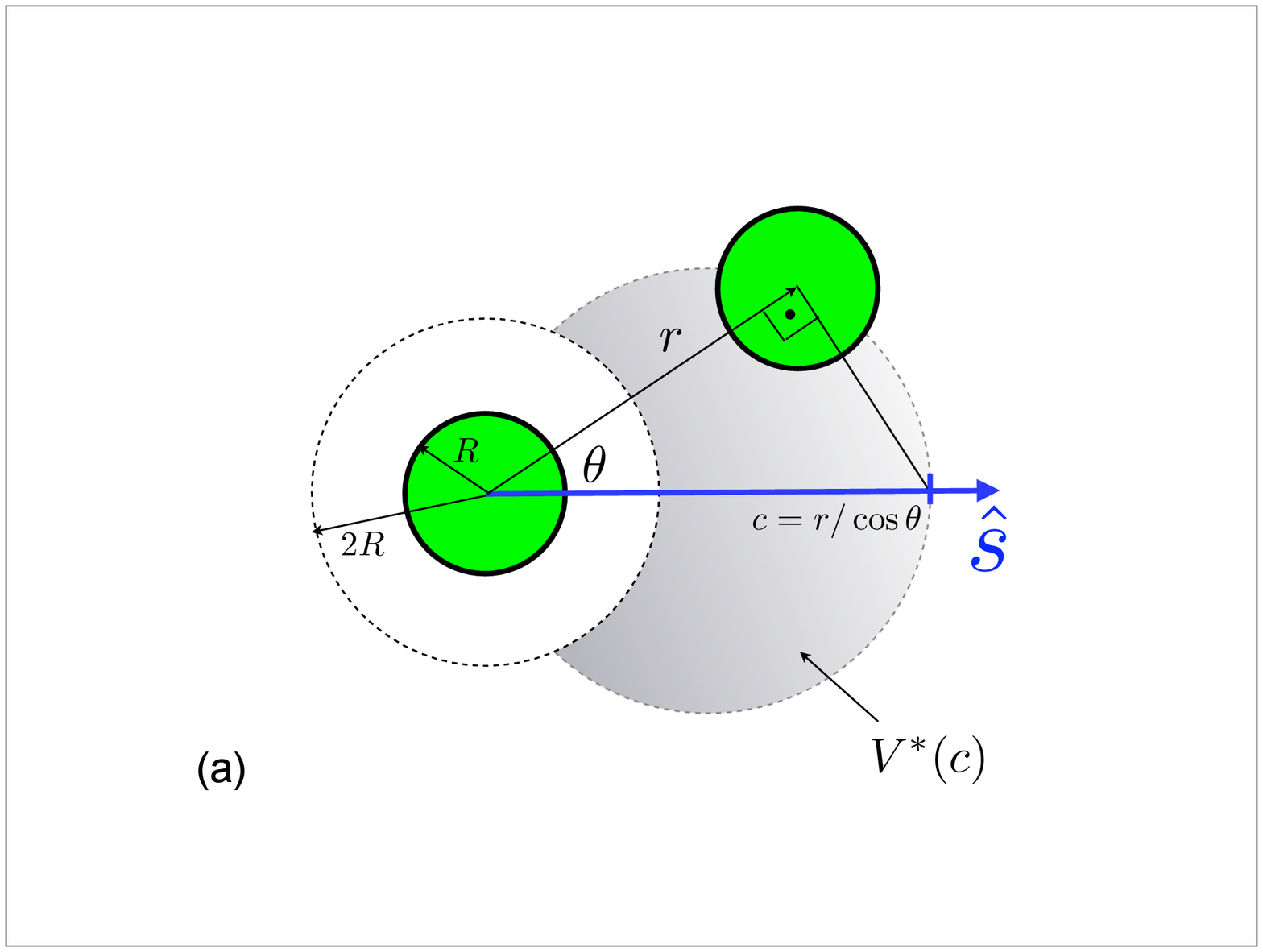}}
\resizebox{8cm}{!}
      { \includegraphics{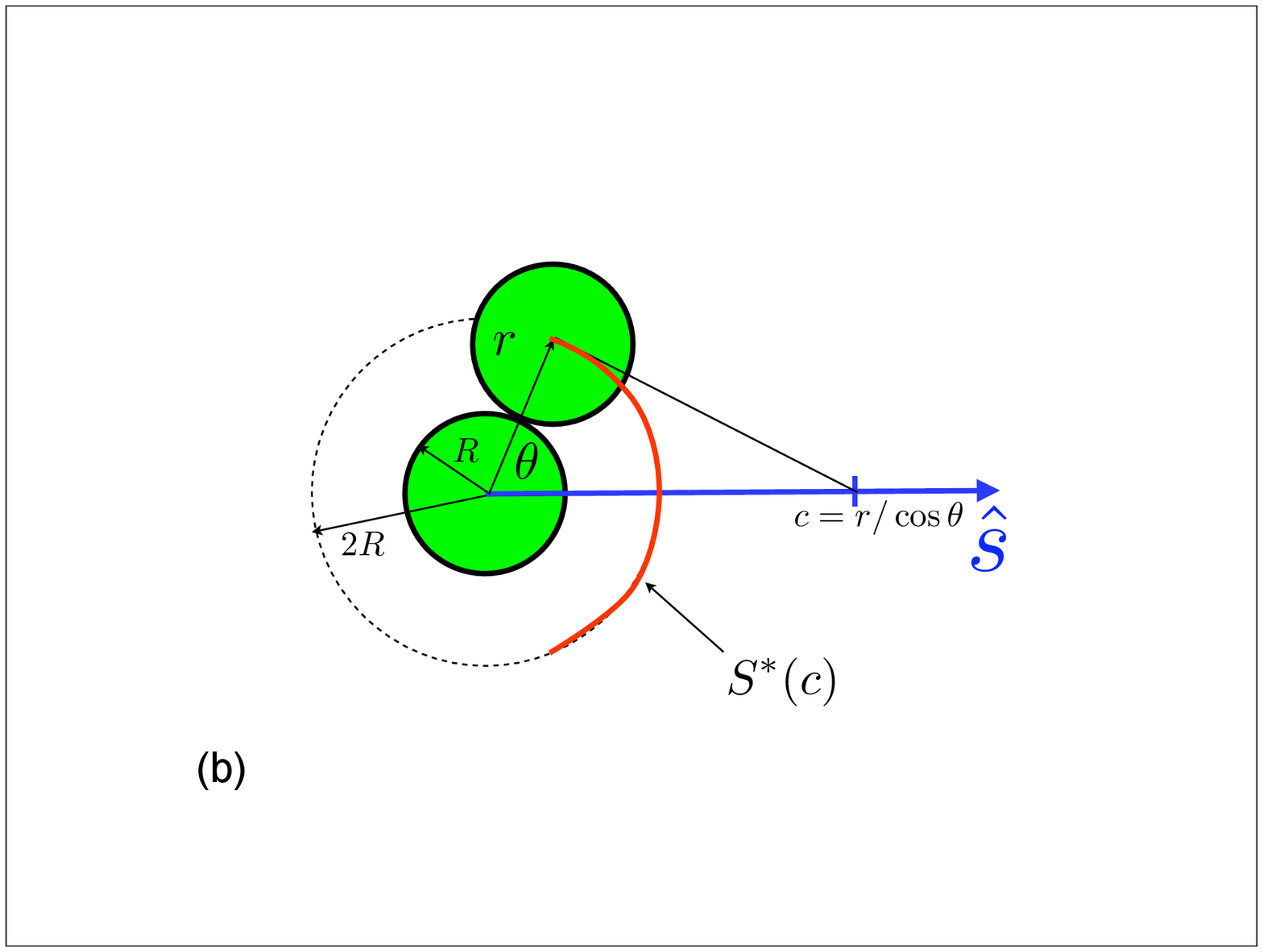}} } }

  \caption{ Schematic illustration of the derivation of $P_{\rm B}$
    and $P_{\rm C}$.  We plot a 2d case for simplicity, although the
    calculation applies to 3d.  (a) $P_{\rm B}$: Background term.
The considered particle (green) is
    located in the center, the closest particle in the $\hat s$
    direction is at $(r,\theta)$, and the white area is the excluded
    zone $r<2R$ for the center of any other grain. For a fixed $c$,
    the grey area with volume $V^*(c)$
is the region of the plane $(r',\theta')$ where $r'
    / \cos \theta' < c $.  If the particle at $(r,\theta)$ is the
    Voronoi particle defining the boundary of the Voronoi cell in the
    $\s$ direction, then no other particle is in the grey zone in the
    figure.  The computation of $P_{\rm B}$ involves the calculation of the
    probability that all the remaining particles in the packing are
    outside the grey excluded volume, $V^*(c)$.
(b) $P_{\rm C}$: Contact term. The calculation of this term involves the
probability to find all the contact particles away from the red area defined
by the closest contact particle to the direction $\s$.
}
\label{volumec}
\end{figure}

Next, we develop a statistical theory for the probability to find a
Voronoi volume in order to calculate the mesoscopic volume function by
averaging the single grain function like in Eq. (\ref{phi}).  For a
given grain, $i$, the calculation reduces to finding the ball $j$ with
$\min_j r_j/\cos\theta_j$ where the minimization is over all grains
(see Fig. \ref{volumec} for notation). We consider $r_j=r_{ij}$,
$\cos\theta_j= \s\cdot\drij$, and we set $2R=1$ for simplicity.  While
the form Eq. (\ref{phi}) is valid for any dimension, the following
statistical theory differs for each dimension.  In what follows we
work in 3d.  Results in 2d and large dimension will be presented in
future papers.

We consider the minimal particle contributing to the Voronoi volume
along the $\s$ direction at $(r,\theta)$ (see Fig. \ref{volumec}) with

\begin{equation}
c=r/\cos\theta.
\end{equation}
We call this particle, the Voronoi particle.  Then the quantity to
compute is the probability to find all the remaining particles in the
packing at a distance larger than $c$ from particle $i$ along the $\s$
direction. That is, we compute the inverse cumulative distribution
function, denoted $P_>(c)$, to find all the grains $j$ with
$r_j/\cos\theta_j>c$, and therefore not contributing to the Voronoi
volume.

In terms of Fig. \ref{volumec}a, the minimal Voronoi particle
located at $c$, determining the Voronoi cell, defines an excluded
volume, represented by the gray volume in the figure, where no
other particle can be located (otherwise they would contribute to
the Voronoi volume). We denote this excluded volume as $V^*(c)$.
Thus, $P_>(c)$ represents the probability that the remaining
particles in the packing have $r_j/\cos\theta_j$ larger than $c$
and therefore are outside the gray volume $V^*(c)$ in Fig.
\ref{volumec}a.

If we know this inverse cumulative distribution, then the mesoscopic
free volume function is  obtained as the mean value of $w_i^s$
over the probability density $-\frac{dP_>(c)}{dc}$:
\begin{equation}
w \equiv
\langle w_i^s\rangle_i = \langle c^3\rangle-1,
\end{equation}
and
\begin{equation}
\begin{split}
  w \equiv \langle w_i^s\rangle_i =  \int_1^\infty (c^3-1)
  \frac{d[1-P_>(c)]}{dc} dc =\\ -
\int_1^\infty (c^3-1)
  \frac{dP_>(c)}{dc} dc = \\
 -\int_1^0 (c^3-1) dP_>=
\int_0^1 (c^3-1) dP_>.
\end{split}
\label{phi2}
\end{equation}

The integration in Eq. (\ref{phi2}) ranges from $1$ to $\infty$
with respect to $c$ since the minimum distance for a ball is for
$r=1$ and $\theta=0$ giving $c=1$ and the maximum at $r\to
\infty$. When changing variables to $dP_>$, the limits of
integration $c:[1,\infty)$ correspond to the inverse cumulative
distribution function $P_>:[1,0]$.

Next, we calculate the inverse cumulative distribution $P_>(c)$.
Considering the Voronoi particle at distance $c$, the remaining
balls are in the bulk and in contact with the ball $i$.  A crucial
step in the calculation is to separate the distribution into two
contributions: a contact term, $P_{\rm C}(c)$, and a bulk term,
$P_{\rm
  B}(c)$.
This separation is important because it allows one to obtain the
dependence of the average volume function on the coordination
number. The contact term naturally depends on the number of
contacting particles $z$, while the bulk term depends on the
average $w$.  Since $P_>(c)$ represents the probability that all
the balls in the packing except the Voronoi ball are outside the
grey volume in Fig. \ref{volumec}a, then the geometrical
interpretation of the contact and bulk term is the following:
\begin{itemize}
\item $P_{\rm B}(c)$ represents the probability that the balls in the
  bulk are located outside the grey volume $V^*(c)$ in
  Fig. \ref{volumec}a.
\item $P_{\rm C}(c)$ represents the probability that the contact balls
  are located outside the boundary of the grey area marked in
  red in Fig. \ref{volumec}b and denoted $S^*(c)$.
\end{itemize}

An assumption of the theory is that both contributions are
considered independent. Therefore:

\begin{equation}
  P_>(c) =
  P_{\rm B}(c) P_{\rm C}(c).
  \label{indep}
\end{equation}
Notice that the meaning of Eq. (\ref{indep}) is the following: we
first assume that the Voronoi particle is located at $c$. Then all
remaining particles, including contact and bulk particles, are
outside the grey excluded zone. This results in the multiplication
of the probabilities as in Eq. (\ref{indep}).

\subsection{Calculation of  $ P_{\rm B}(c)$ and $ P_{\rm C}(c)$}

In order to calculate the distributions $P_{\rm B}$ and $P_{\rm C}$ we
apply the following assumptions.

\begin{figure}
  \centering {
  \resizebox{8cm}{!} { \includegraphics{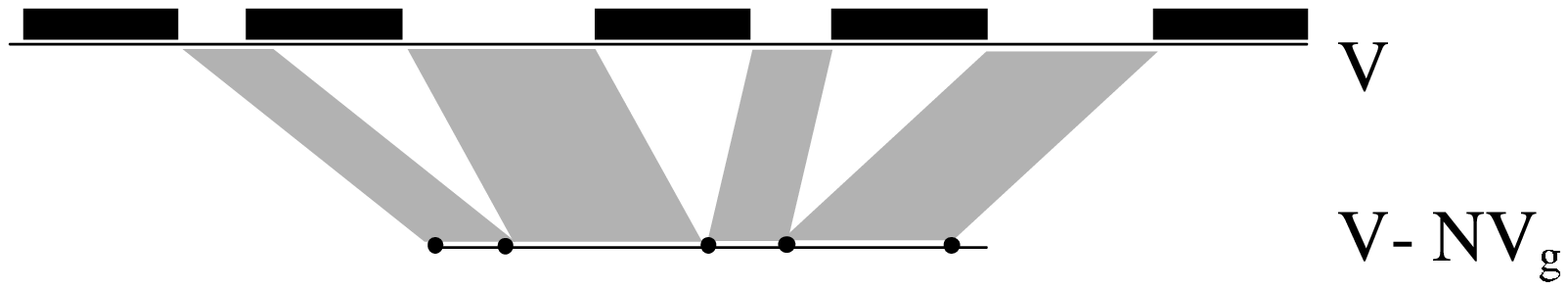}}} \caption {A
    mapping between hard sphere and ideal gas in the one dimensional
    system. A system of total volume $V$ of $N$ rods in 1d (hard
    spheres in 3d with volume $V_g$) can be mapped to a system of $N$
    points of total size $V-NV_g$ by simply removing the size of the
    balls. This mapping allows to calculate $P_>$ exactly in 1d which
    is shown to be an exponential as in Eq. (\ref{expo}) in the large
    $N$ limit. For higher dimensions, we cannot just remove the size
    of the balls $NV_g$ since the shape of the balls is important as
    well. This implies that the exponential probability is an
    approximation to the real distribution in 3d, as discussed in the
    text.} \label{1d_map}
\end{figure}

We treat the general case of calculating the probability for $N$
particles in a system of volume $V$ to be outside a given volume
$V^*$ when added at random.  In 3d, this probability is nontrivial
since the volume occupied by each ball in the packing should be
greater than the size of the ball. However, this probability can
be calculated exactly in the case of 1d in the large $N$ limit. In
1d, the distribution of possible arrangements of hard-rods
corresponds to the distribution of ideal gas particles by removing
the volume occupied by the size of the ball as we can see in Fig.
\ref{1d_map}.  Such a mapping is exact in 1d, implying an
exponential distribution of the free volume.

The probability to locate one particle at random outside the volume
$V^*$ in a system of volume $V$ is $$P_>(1)=(1-V^*/V).$$ For $N$
independent particles, we obtain:
$$
P_>(N)=(1-V^*/V)^N.
$$
We set $V^*/V=1/x$ and the particle density $\rho=N/V$. Then
\begin{equation}
P_>(N) = (1-1/x)^{\rho V}
= (1-1/x)^{\rho x V^*}.
\end{equation}
In the limit of a large number of particles, $x \rightarrow \infty$,
we obtain a Boltzmann-like exponential distribution for the probability
of $N$ particles to be outside a volume $V^*$:
\begin{equation}
P_>(N) \propto \exp(-
  \rho V^*), \,\,\, N\to \infty.
\label{expo}
\end{equation}

While the above derivation is exact in 1d, the extension to higher
dimensions is an approximation, since there exist additional
geometrical constraints. Even if there is a void with enough volume to
be occupied by a particle (the volume of the void is larger or equal
than the size of the particle), the constraint imposed by the
geometrical shape of the particle (which does not exist in 1d) might
prevent the void from being occupied.  Nevertheless, in what follows,
we assume the exponential distribution of Eq. (\ref{expo}) to be valid
in 3d as well.

The background is assumed to be uniform with a mean free bulk
particle density given by:

\begin{equation}
\rho(w) = \frac{N}{V- N V_g}= \frac{N}{N V_g \phi^{-1}- N V_g} = \frac{1}{V_g w},
\label{rho}
\end{equation}
where we define the volume fraction $\phi^{-1} =w + 1$. Therefore,
$P_{\rm B}$ assumes a Boltzmann-like distribution of the form
analogous to Eq. (\ref{expo}),
\begin{equation}
  P_{\rm B}(c) = \exp\Big(-\rho(w) V^*(c)\Big),
\label{expob}
\end{equation}
where

\begin{equation}
\begin{split}
  V^*(c)& = 2\pi
  \int\Theta(c-r/\cos\theta) d\vec{r}\\
  &=2\pi
  \int_{1}^{\infty}\int_{0}^{\pi/2}\Theta(c-r/\cos\theta)r^2\sin\theta
  d\theta dr \\
  &=2\pi\int_{1}^{c}r^2\int_{0}^{\arccos(r/c)}\sin\theta d\theta dr \\
  &=2\pi\int_{1}^{c}(1-r/c)r^2dr
\end{split}
\end{equation}
is the volume of the grey area in Fig. \ref{volumec}a.  We obtain:
\begin{equation}
V^*(c) = V_g\left((c^3-1)-3(1-\frac{1}{c})\right).
\end{equation}
Therefore,
\begin{equation}
P_{\rm B}(c) =\exp\Big[-\frac{(c^3-1)-3(1-1/c)}{w}\Big].
\end{equation}

The derivation of the surface term, $P_{\rm C}(c)$, is analogous to
that of the volume term.
$P_C(c)$ is assumed to have the same exponential form, analogous to
the background form of $P_B(c)$, Eq. (\ref{expob}), despite not having
the large number approximation leading to Eq.  (\ref{expo}):

\begin{equation}
  P_{\rm C}(c) = \exp\Big(-\rho_S(z) S^*(c)\Big),
  \label{pc}
\end{equation}
where
\begin{equation}
  S^*(c) = 2\pi \int_{0}^{\arccos(1/c)}\sin\theta d\theta = 2\pi(1-\frac{1}{c}),
  \label{sc}
\end{equation}
is the excluded area marked in red in Fig. \ref{volumec}b.

To define $\rho_S(z)$ we first follow the analogy with the bulk
density Eq. (\ref{rho}), $\rho(w)$ to obtain the mean free
surface particle density on the sphere with $z$ contacting
particles:

\begin{equation}
  \rho_S(z) = \frac{z}{4\pi - zS_{\rm occ}},
\label{Sf}
\end{equation}
where $S_{\rm occ} = 2\pi\int_0^{\pi/6}\sin\theta d\theta = 2\pi a$ is
the surface occupied by a single contact ball, with $a = 1-\sqrt{3}/2$
a small value.

However, we notice that the analogy with Eq. (\ref{expo}) is more
difficult to justify here since the large number limit is lacking for
the surface term: the maximum number of contacting spheres, the
so-called kissing number, is 12 and for disordered packings is 6 in
average. Thus, rather than considering Eq. (\ref{pc}) as the exact
form of the surface term, we take the exponential form as a simple
variational Ansatz,
where the surface density $\rho_S(z)$ is the variational parameter
which has to be corrected from Eq. (\ref{Sf}) due to the small number
of the balls on the surface.





\begin{figure}
  \centering {
      \resizebox{6.5cm}{!}
      {\includegraphics{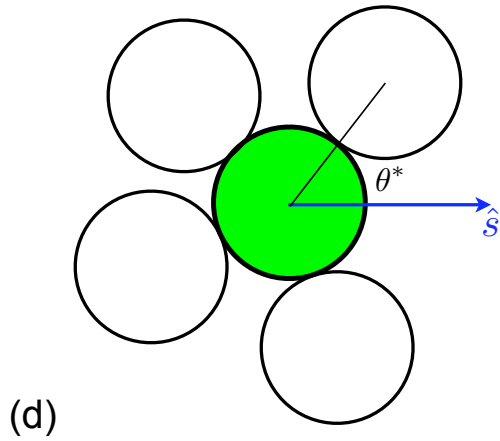}} }
\caption{
The calculation of the surface term $P_{\rm C}$ involves the
mean free surface density for a given $z$ (=4 in this example) obtained from the
angle $\theta^*$. Note that we show the 2d case for simplicity, but the
$\theta^*$ corresponds to a solid angle in 3d.
}
\label{volumed}
\end{figure}

A more physical definition of $\rho_S(z)$ is a mean free density
representing the inverse of the average of $S^*(c)$: $\langle S^*(c)
\rangle$.  The meaning of $\langle S^*(c) \rangle$ for a given number
of contacting particles is the following: add $z$ contact particles at
random around a central sphere.  The average of the solid angles of
the gaps left between nearest neighbor contacting spheres is $\langle
S^*(c) \rangle$ (see Fig. \ref{volumed}).  Indeed we obtain:

\begin{equation}
\begin{split}
\langle S^*\rangle &= \int_1^\infty S^*(c)\frac{d[1-P_C(c)]}{dc}dc \\
&=  \int_1^\infty S^*(c)\rho_S\exp(-\rho_S S^*(c))\frac{dS^*}{dc}dc \\
&=  \rho_S\int_0^{2\pi} S^*\exp(-\rho_S S^*)dS^* \\
&\approx  \rho_S\int_0^{\infty} S^*\exp(-\rho_S S^*)dS^* \\
&= 1/\rho_S
\end{split}
\end{equation}
Then, the surface density is replaced by the following definition:
\begin{equation}
  \rho_S = \frac{1}{\left<S^*\right>}
\end{equation}

Under these considerations, in order to estimate the value of
$\rho_S(z)$ we first consider a single particle approximation.  We
calculate the mean of $S^*$ for a single particle $z = 1$, which gives
$\left<S^*\right>=2\pi$, since $S^*$ ranges uniformly from $0$ to
$4\pi$.  We note that the value is different from the prediction of
Eq. (\ref{Sf}) in the large number limit. Since we expect $\rho_S(z)$
to be proportional to $z$, under a single particle approximation we
find:

\begin{equation}
 \rho_S(z) \approx
\frac{z}{2\pi}, \,\,\,\,\,\, \mbox{single particle approximation}.
\label{single}
\end{equation}

\begin{figure}
  \centering {
  \resizebox{8cm}{!} { \includegraphics{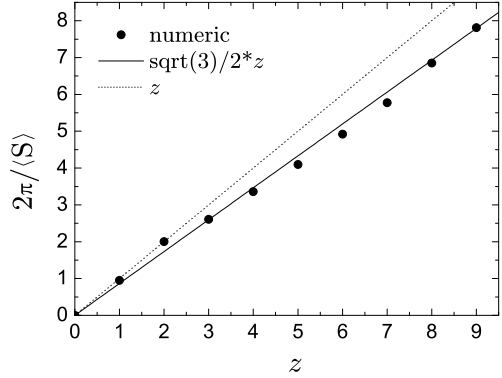}}} \caption {A
    numerical calculation using packing of spheres confirms that $2\pi
    \rho_S(z) = 2\pi/\langle S^*\rangle$ (dots) slightly deviates from
    the single particle estimation as $z$ (dash line), and is much
    better approximated by $(\sqrt{3}/2)z$ with particle size
    correction (solid line).}
\label{sqrt}
\end{figure}

Simulations considering many $z$ contacting particles suggest that
corrections from the single particle approximation are important
for larger $z$. Indeed, a more precise value is obtained from
simulations:

\begin{equation}
  \rho_S(z) \approx \frac{z}{2\pi} \frac{\sqrt{3}}{2}, \,\,\, z>1,
\label{many}
\end{equation}
deviating from the single particle approximation of Eq.
(\ref{single}). This numerical calculation consists of adding $z$
random, non-overlapping, equal-size spheres at the surface of a
ball. For every fix $z$, the sphere closest to the direction $\s$
defines the free angle $\theta^*$ and the surface $S^*=2\theta^*$
(see Fig. \ref{volumed}).  Thus, the idea is to measure the mean
gap, $\left<S^*\right>$, left by $z$ contacting particles along
the $\s$ direction. As seen in Fig. \ref{sqrt}, simulations show
that the more precise value, Eq. (\ref{many}) is valid
rather than the single particle approximation, Eq. (\ref{single}).




This many-particle result can be explained by adding a small
correction term from the area occupied by one particle into Eq.
(\ref{single}).
In the case of many contact particles, $z>1$, many body constraints
imply that the surface, $\left<S^*\right>$, should be corrected by the
solid angle extended by a single ball, and at the same time reduced by
the increasing number of contacting particles.  Thus, up to first
order approximation, $$\left<S^*\right> \approx (2\pi + S_{\rm
  occ})/z.$$ This analysis provides a correction to the surface term
which can be approximated as
\begin{equation}
\begin{split}
  \rho_S(z) = &\frac{z}{2\pi + S_{\rm occ}} = \frac{z}{2\pi (1+a)}  \\
  \approx & \frac{z}{2\pi} (1-a) = \frac{z}{2\pi} \frac{\sqrt{3}}{2},
  \,\,\, z>1,
\end{split}
\label{sqrt3}
\end{equation}
It is clear that the above derivation is by no means exact. It
merely interprets the origin of the correction from the single
particle approximation in order to obtain a proper estimation of
the surface density.  The obtained value agrees very well with the
computer simulations results of Fig. \ref{sqrt} and Eq.
(\ref{many}), and therefore we use it to define $P_{\rm C}(c)$. In
comparison with numerical simulations, the derived form of $P_{\rm
C}(c)$ compares well as we will show in Fig. \ref{pbpc}.
Furthermore, the predicted average volume fraction compares well
with experiments on random close packings.




From Eqs. (\ref{pc}), (\ref{sc}) and (\ref{sqrt3}), we obtain the
surface term:
\begin{equation}
  P_\mathrm{C}(c) = \exp\Big[-\sqrt{3}z(1-1/c)/2\Big],
\end{equation}
and the inverse cumulative distribution of volumes takes the form:
\begin{equation}
  P_>(c)> =
  \exp\left[-\frac{1}{w}\left((c^3-1)-3(1-\frac{1}{c})\right)-
    \frac{\sqrt{3}}{2}z(1-\frac{1}{c})\right]. \label{pc1}
\end{equation}

\section{The mesoscopic volume function}
\label{meso}

Substituting Eq. (\ref{pc1}) into Eq. (\ref{phi2}), we obtain a
self-consistent equation to calculate the average free volume, $w$:

\begin{equation}
\begin{split}
  w = \int_0^1 (c^3-1) \,\,\,d
  \exp \Big[-\frac{1}{w}\Big((c^3-1)-3(1-\frac{1}{c})\Big)- \\
    \frac{\sqrt{3}}{2}z(1-\frac{1}{c})\Big].
\end{split}
\label{w1}
\end{equation}

Since

\begin{equation}
w = \int_0^1 (c^3-1) d \exp[-(c^3-1)/w],
\end{equation}
then Eq. (\ref{w1}) can be solved exactly, leading to an analytical
form of the free average volume function, under the approximations of
the theory. The fact that we can solve this equation exactly is a
fortuitous event.  It is worth mentioning that an analogous analysis
performed in 2d as well as in infinite dimensions does not lead to an
analytical solution of the self-consistent equation (\ref{w1}) and
therefore only numerical solutions are possible for the average volume
function in those dimensions under similar assumptions as used in 3d.

To solve Eq. (\ref{w1}) we start from the identity:
\begin{equation}
  \int_{0}^{\infty}
  \frac{x}{w} \exp(-\frac{x}{w})dx = 1.
\end{equation}
Then we find:
\begin{equation}
\begin{split}
1 = \int_{0}^{1} \frac{1}{w}\Big((c^3 - 1) - \alpha
(1-\frac{1}{c})\Big) \times \\
d \exp\left[-\frac{1}{w}\left((c^3-1)-
\alpha (1-\frac{1}{c})\right)\right],
\end{split}
\end{equation}
where $\alpha = 3 - wz \sqrt{3}/2$. Or
\begin{equation}
\begin{split}
  0 & = \int_{0}^{1} \frac{1}{w} (c^3 - 1) d
  \exp\left[-\frac{1}{w}\left((c^3-1)- \alpha
      (1-\frac{1}{c})\right)\right] - 1 \\
  & = \alpha \int_{0}^{1} \frac{1}{w} \left(1-\frac{1}{c}\right) d
  \exp\left[-\frac{1}{w}\left((c^3-1)- \alpha
      (1-\frac{1}{c})\right)\right].
\end{split}
\label{w2}
\end{equation}

The second integration in the right hand side is equal to zero
only at $w = 0$ or $w \rightarrow \infty$, corresponding to two
trivial solutions at $\phi = 1$ and $\phi = 0$, respectively. The
only non-trivial solution happens at $\alpha = 0$, and therefore
we find
\begin{equation}
  w(z)= \frac{2\sqrt{3}}{z} \,\,\,\,
  \Rightarrow \,\,\,\, W(z) = \frac{2\sqrt{3}}{z} V_g.
\label{phi3}
\end{equation}

\begin{figure}
\centering {
\vbox {
 \resizebox{7.5cm}{!}{\includegraphics{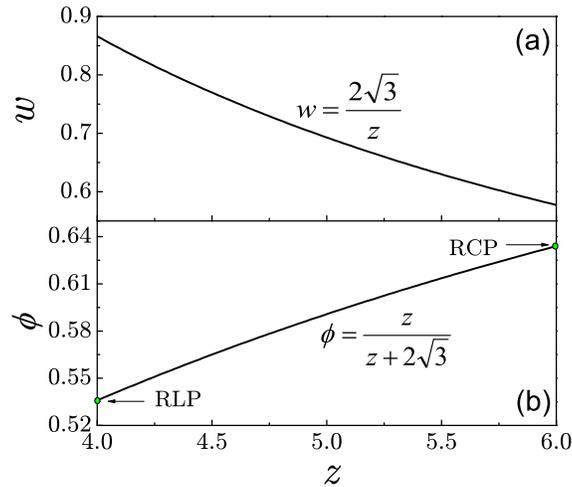}}
}
}
\caption{ (a) Mesoscopic volume function of granular matter, $w(z)$
versus $z$. (b) Equation of state relating the volume fraction and
coordination number in the limit of infinite compactivity. }
\label{state}
\end{figure}

We arrive at a mesoscopic volume function (plotted in
Fig. \ref{state}a) which is more amenable to a statistical mechanics
approach for jammed matter.  Equation (\ref{phi3}) is a coarse-grained
``Hamiltonian'' or volume function that replaces the microscopic
Eq. (\ref{vor1}) to describe the mesoscopic states of jammed
matter. While Eq. (\ref{vor1}) is difficult to treat analytically in
statistical mechanics, the advantage of the mesoscopic
Eq. (\ref{phi3}) is that it can easily be incorporated into a
partition function since it depends only on $z$ instead of all degrees
of freedom ${\vec r}_i$.


If the system is fully random, so that we can extend the assumption of
uniformity from the mesoscopic scales to the macroscopic scales, we
arrive to an equation of state relating $\phi^{-1}= w + 1$ with $z$
as:
\begin{equation}
\phi=\frac{z}{z+ 2\sqrt{3}}.
\label{phi4}
\end{equation}


Thus, Eq. (\ref{phi4}), plotted in Fig. \ref{state}b, can be
interpreted as a equation of state for fully random jammed matter.
We recall that it has been obtained under the approximation of a
packing achieving a completely random state; the proper test of
such an equation would be to add spheres at random and observe
their distribution and average values.
In a second paper of this series \cite{jamming2} we will show that it
corresponds to the equation of state in the limit of infinite
compactivity when the system is fully randomized.  Indeed, Eq.
(\ref{phi4}) already provides the density of random loose packings and
random close packings when the coordination number is replaced by the
isostatic limits of 4 and 6, for infinitely rough and frictionless
particles, respectively.

So far, our analysis only includes geometrical constraints, but
has not made use of the mechanical constraint of jamming.  The
jamming condition can be introduced through the condition of
isostaticity which applies to the mechanical coordination number,
$Z$, for rigid spherical grains \cite{alexander}.  For isostatic
packings the mechanical coordination number is bounded by
$Z=d+1=4$ and $Z=2d=6$ and we obtain the minimum and maximum
volume fraction from Eq. (\ref{phi4}) as $\phi_{\rm RLP} = 4/(4+2
\sqrt{3}) \approx 0.536$ and $\phi_{\rm RCP} = 6/(6+2 \sqrt{3})
\approx 0.634$ as shown in Fig. \ref{state}b. We identify these
limits as the random close packing (RCP) and random loose packing
(RLP) fractions (the maximum and minimum possible volumes of
random packings of spherical particles).  This result is fully
explored in Jamming II \cite{jamming2} where the phase space of
jammed configurations is obtained using Edwards statistical
mechanics of jammed matter based on Eq. (\ref{phi3}). We note
that the mechanical coordination, $Z$, is not the same as the
geometrical one, $z$ (see \cite{jamming2} for details), and an
equation of state of jammed matter should be a relation between
state variables, such as average global volume fraction $\phi$ and
mechanical coordination number $Z$. However, it can be shown that
in the limit of infinite compactivity, the mechanical coordination
number $Z$ is identical to the geometrical coordination number
$z$, and the equation of state in this case has the same form of
Eq. (\ref{phi4}). Thus for now, it suffices to say that we can
obtain the two packing limits from Eq. (\ref{phi4}) in the case of
fully random systems of infinite compactivity.



The experimental studies of \cite{aste} on the free volume versus
coordination number of grains in tomography studies of random packings
of spheres support Eq. (\ref{phi4}) (see Fig. 6 in \cite{aste}).

\subsection{Quasiparticles}
\label{quasi}


Equation (\ref{phi3}) should be interpreted as representing
quasiparticles with free volume $w$ and coordination number $z$.  When
a grain jams in a packing, it interacts with other grains.  The role
of this interaction is assumed in the calculation of the volume
function (\ref{phi3}) and is implicit in the coarse-graining procedure
explained above.  Thus, the quasiparticles can be considered as
particles in a self-consistent field of surrounding jammed matter.  In
the presence of this field, the volume of the quasiparticles depends
on the surrounding particles, as expressed in Eq. (\ref{phi3}).
The assembly of quasiparticles can be regarded as a set of
non-interacting particles (when the number of elementary excitations
is sufficiently low).  The jammed system can be considered as an ideal
gas of quasiparticles and a single particle partition function can be
used to evaluate the ensemble. These ideas are exploited in the
definition of the partition function leading to the solution of the
phase diagram discussed in Jamming II \cite{jamming2}.


\section{Numerical Tests}
\label{test}

In this section, we wish to test some predictions and approximations
of the theory and propose the necessary improvements where needed.
The purpose of this calculation is, first, to evaluate the predictions
of the theory regarding the inverse cumulative volume distributions,
and, second, to test whether the background distribution is
independent of the contact distribution by comparing $P_{\rm B}(c)
\times P_{\rm C}(c)$ with $P_{>}(c)$.

It is important to note that the predictions of the mesoscopic
theory refer to quasiparticles as discussed above. The
quasiparticles of free volume Eq. (\ref{phi3}) of fix coordination
number can be considered as the building blocks or elementary
units of jammed matter.  To properly study their properties and
test the predictions and approximations of the present theory one
should in principle isolate the behavior of the quasiparticles and
study their statistical properties such as the distribution of
volumes and their mean value as a function of coordination number.
Such a study is being carried out and may lead to a more precise
solutions than the one presented in the present paper.

Nevertheless, in what follows we take an approximate numerical route
and study the behavior of quasiparticles directly from computer
generated packings with Molecular Dynamics. While such packings
already contain the ensemble average of the quasiparticle, we argue
that in some limits they could provide statistics for isolated
quasiparticles, at least approximately.  This is due to a result that
needs to wait until Jamming II \cite{jamming2}, where we find that for
a system of frictionless packings there is a unique state of jamming
at the mesoscopic level and therefore the compactivity and the
ensemble average does not play a role, at least under the mesoscopic
approximation developed here.  Therefore, below, we use frictionless
packings in order to test the theory. We note, though, that the
conclusions of this section remain approximative, waiting for a more
precise study of the packings of fixed coordination number.

We prepare a frictionless packing at the jamming transition with
methods explained in Jamming II.  The packing consists of 10,000
spherical particles interacting with Hertz forces.  In the simulation,
we pick up a direction $\hat{s}$ randomly, and collect $c_b$ and $c_c$
from the balls, as follows:



\begin{eqnarray}
c_b &=& \min_{\hat{s}\cdot\hat{r}_{ij} >
0}\frac{r_{ij}}{2\hat{s}\cdot\hat{r}_{ij}}, r_{ij} > 2R, \\
\nonumber c_c &=& \min_{\hat{s}\cdot\hat{r}_{ij} >
0}\frac{r_{ij}}{2\hat{s}\cdot\hat{r}_{ij}}, r_{ij} \leq 2R.
\end{eqnarray}


Thus, for a given direction, we find two minimum values of $c$
independently as $c_b$ and $c_c$ over all particles in the
packing.  We then collect data for 400 different directions.  The
$c_b$ is only provided by the balls in the background, and $c_c$
is only provided by the balls in contact.  From the probability
density, we then calculate the cumulative probability to find a
ball at position $r$ and $\theta$ such that $r/\cos \theta > c$.
That is, we calculate the cumulative distribution of $c_b$ and
$c_c$ individually, i.e., $P_{\rm B}(c_b)$ and $P_{\rm C}(c_c)$.

\begin{figure}
\centering {
\vbox {  \resizebox{7cm}{!}{\includegraphics{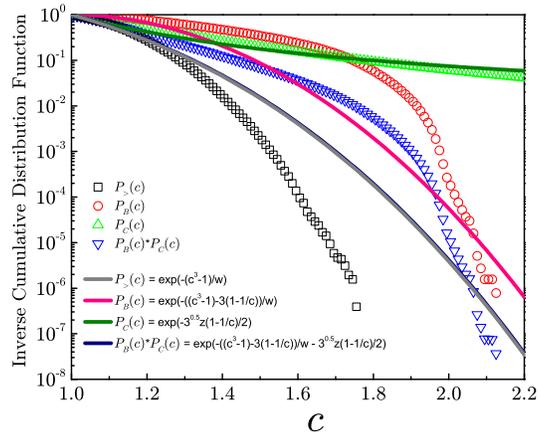}}
}}
\caption{Comparison between theory and simulations for the inverse
  cumulative distributions, $P_{\rm B}(c), P_{\rm C}(c)$, $P_{\rm
    B}(c) \times P_{\rm C}(c)$ and $P_{>}(c)$ for a packing at the
  frictionless point with $z=6$ as explained in Jamming II
  \cite{jamming2}.  The choice of a frictionless packing to test the
  distribution of volumes is due to the fact that these packings are
  independent of the compactivity as will be shown in Jamming II. In
  general such an equation for the distribution of $c$ should be
  tested by generating local packings by randomly adding $z$ balls
  surrounding a given sphere.  }
\label{pbpc}
\end{figure}

The inverse cumulative distributions are plotted Fig. \ref{pbpc}
showing that the theory approximately captures the trend of these
functions but deviations exist as well, especially for $c$ values
larger than its average. The contact term $P_{\rm C}(c)$ is well
approximated by the theory supporting the approximations involved in
the calculation of the surface density term, Eq. (\ref{sqrt3}).  The
background term shows deviations for larger $c$; for smaller $c$ the
theory is not too far from simulations. From Fig. \ref{pbpc} we notice
that $P_{>}(c)$ is well reproduced by the theory up to a value of 10\%
of its peak value.

It is important to note that the mesoscopic volume function, $w$, is
extracted from the mean value of $\langle w^s\rangle$ as $w=\langle
w^s\rangle= \langle c^3\rangle -1 $. While some deviations are found
in the inverse cumulative distribution, we find that the average value
of the volumes are well approximated by the theory.  Indeed, we find
that the deviations from the theoretical probabilities for
$w^s>\langle w^s\rangle $ appear not to contribute significantly
towards the average volume function.

For instance, the packing in Fig. \ref{pbpc} has a volume fraction
$\phi=0.64$ as measured from the particle positions. This value
agrees with the average $\langle w^s\rangle$ obtained from the
prediction of the probability distribution $P_>(c)$.  We find
$\langle c^3\rangle$ = 1.561, then $\langle w^s\rangle$ = $\langle
c^3\rangle -1 = 0.561$ and $\phi = 1/\langle c^3\rangle =
1/(\langle w^s\rangle+1) = 0.641 $ in agreement with the volume
fraction of the entire packing obtained from the position of all
the balls, $0.64$.

Thus, the present theory gives a good approximation to the average
Voronoi volume needed for the mesoscopic volume function, even though
the full distribution presents deviations from the theory.  In order
to capture all the moments of the distribution a more refined theory
is needed.  Such a theory will include the corrections to the
exponential forms of $P_{\rm B}(c)$ and $P_{\rm C}(c)$ and their
correlations.  The main result of the mesoscopic theory, being the
average Voronoi volume decreasing with the number of contacts, is not
affected by the assumptions of the theory for the full probability
distribution.

The correlations between the contact and bulk term are quantified
by comparing $P_{\rm B}(c) \times P_{\rm C}(c)$ with $P_{>}(c)$ in
Fig. \ref{pbpc}.  From the figure we see that below and around the
mean $\langle c \rangle$, the full distribution is close to the
theoretical result while deviations appear for larger $c$. Further
testing of the existence of correlations between $P_{\rm B}(c)$
and $P_{\rm C}(c)$ is obtained by calculating the product-moment
coefficient of Pearson's correlation as follows.



The Pearson's coefficient is:

\begin{equation}
r^2=\frac{{S_{bc}}^2}{S_{bb}S_{cc}},
\end{equation}
where $S_{bb} = \Sigma({c_b}^2-{{\bar c}_b}^2)$, $S_{cc} =
\Sigma({c_c}^2-{\bar{c}_c}^2)$, and $S_{cb} =
\Sigma(c_bc_c-\bar{c}_b\bar{c}_c)$. We find that the Pearson
coefficient $r^2=0.0173$ is close to zero, meaning that the
correlations between $P_B(c)$ and $P_C(c)$ are weak.

The present numerical results imply that the current assumptions
of the theory are reasonable. The conclusions are that while the
cumulative distributions present deviations from the theory in
their tails, the average value of the Voronoi volumes are well
captured by the approximations of the theory, therefore providing
an accurate value for the volume function.

More importantly, the present approach indicates a way to improve
the theory to provide more accurate results.
Our current studies indicate that an exact solution of the
distribution, $P_>(c)$, may be possible up to the second
coordination shell of particles, for a fixed z-ensemble.  Due to
the fact that the range of Voronoi cell is finite, it is possible
to work out a description for the finite, but large, number of
degrees of freedom for both disordered and ordered packings
through computational linear programming, in principle.  This
approach is related to the Hales' proof of the Kepler conjecture
\cite{hales}. The present theory is a mean-field version in terms
of the restricted description of the disordered packings, which
allows us to reduce the dimensionality of the original problem in
order to write down the analytic form of the volume function in
reasonable agreement with known values of RCP and RLP.
The present approximations of the theory are further supported by
agreement between the obtained form of the volume function and the
empirical findings of the experiments of \cite{aste}.  In a future
paper we will present a more exact theory of the volume fluctuations
capturing not only the mean value but also higher moments.



It is of interest to test the formula Eq. (\ref{phi3}) with the
well-known example of the FCC lattice at $z=12$ to assess the
approximations of the theory.  At this limiting number of neighbors
the entire class of attainable orientational Voronoi cells have
volumes in a very narrow range around $0.7V_g$ which is larger than
the prediction from Eq. (\ref{phi3}).  The free volume of the FCC
Voronoi cell is $0.35135$ while the mesoscopic volume function for
$z=12$ gives $2\sqrt{3}/12 = 0.2886$, below the real value.  We
explain this discrepancy since the current theory is developed under
the assumption of isotropic packings. Isotropic packings are
explicitly taken into account in the theory considering the
orientational Voronoi volume ${\cal W}_i^s$ (along a direction $\hat
s$) as a simplification of the full Voronoi volume, ${\cal W}_i^{\rm
  vor}$. Such a simplification is
meaningful for isotropic packings but fails for anisotropic or ordered
packings.  Indeed, in the case of packings with strong angular
correlations, the ``weak-coupling'' hypothesis employed here does not
work well.  The extension of the current theory to anisotropic
packings, such as the FCC lattice at $z=12$, can be carried out, but
remains outside of the scope of the present work. In this case, the
full Voronoi volume of Eq.  (\ref{vor1}) must be used.

Eventually a volume function that accounts for disordered and ordered
packings is very important.  This volume function could test the
existence of phase transition between the ordered and disordered
phases.
The theory we propose here is a quasiparticle version in terms of
the restricted description of the disordered packings, which
allows us to reduce the dimensionality of the original problem in
order to write down the analytic form of the volume function.  The
fact that Eq. (\ref{phi3}) does not predict correctly the volume
of FCC but does predict correctly the volume of RCP raises the
interesting possibility that there could be a phase transition at
RCP, an intriguing possibility which is being explored at the
moment.

\section{Summary}

\label{outlook}

In summary, we present a plausible volume function describing the
states of jammed matter.  The definition of ${\cal W}_i$ is purely
geometrical and, therefore, can be extended to unjammed systems such
as colloids at lower concentrations.  In its microscopic definition,
the volume function is shown to be the Voronoi volume per particle,
and an analytical formula is derived for it. However this form is
still intractable in a statistical mechanics analysis.  We then
develop a mesoscopic version of ${\cal W}_i$ involving coarse graining
over a few particles that reduces the degrees of freedom to a single
variable, $z$.
This renders the problem within the reach of analytical calculations.
The significance of our results is that they allow the development of
a statistical mechanics to predict the observables by using the volume
function in a Boltzmann-like probability distribution of states
\cite{sirsam} when the analysis is supplemented by the jamming condition
'a la Alexander' \cite{alexander};
which we propose in \cite{jamming2} to be the isostatic condition for
rigid particles.

{\bf Acknowledgements}. This work is supported by NSF-CMMT, and
DOE Geosciences Division. We thank J. Bruji\'c for inspirations
and C. Briscoe for a critical reading of this manuscript.

\end{document}